\documentclass[
reprint,
amsmath,
amssymb,
aip,
apl
]{revtex4-2}

\usepackage{graphicx}
\usepackage{bm}
\usepackage[dvipsnames]{xcolor}
\usepackage{footmisc}

\begin{document}

\makeatletter
\def\frontmatter@thefootnote{%
 \altaffilletter@sw{\@fnsymbol}{\@fnsymbol}{\csname c@\@mpfn\endcsname}%
}%
\makeatother

\graphicspath{{./images}}

\preprint{APS/123-QED}

\title{Supercurrent tuning of the Josephson coupling energy\vspace{10pt}}

\author{Maxwell M. Wisne}
\author{Venkat Chandrasekhar}
\email{Corresponding author\\email: v-chandrasekhar@northwestern.edu}
\affiliation{Department of Physics and Astronomy, Northwestern University, 2145 Sheridan Road, Evanston, IL 60208, USA}

\date{\today}

\begin{abstract}
The ability to non-dissipatively tune the Josephson coupling energy of Josephson junctions is a useful tool in frequency-tunable qubits. This is typically done by threading magnetic flux through two junctions connected in a loop, a geometry that exposes the qubit to magnetic environmental noise. In this paper, we show that by biasing a junction with supercurrent from a separate pair of superconducting leads coupled to the device, the Josephson energy can be tuned without the need for a flux loop. Our multiterminal device may enable the realization of a frequency-tunable qubit with greatly reduced susceptibility to flux noise. \\ 

Keywords: Superconducting qubit, Josephson junction, Josephson energy, critical current, multiterminal, SQUID 
\end{abstract}

\maketitle
The Josephson junction (JJ) serves as the nonlinear inductor element in a superconducting qubit. The qubit frequency is determined by the coupling strength between superconducting electrodes of the JJ, known as the Josephson coupling energy $E_J$. In a superconductor-insulator-superconductor (SIS) JJs, $E_J$ depends on the intrinsic properties of the junction and is directly proportional to the critical current $I_c$, the maximum allowed supercurrent that can flow through the junction, by $E_J = \hbar/2e I_c$. In frequency-tunable qubits, $E_J$ is commonly tuned \textit{in situ} by connecting the two JJs in parallel and threading the device with magnetic flux $\Phi$ induced by field coils driven with current sources integrated into the measurement system.  

Traditionally, superconducting qubits utilize oxide tunnel junctions where the current-phase relation is sinusoidal and $2\pi$-periodic. Recently, however, there has been interest in implementing superconductor-normal metal-superconductor (SNS) JJs as nonlinear quantum circuit elements where the current-phase relation may be non-sinusoidal \cite{dicarlo_2015} and tuned between $\pi$- and $2\pi$-periodicity. \cite{renard_2024} In this case, $E_J$ of the SNS JJ can be modified by tuning the conductivity of the weak link. The qubit frequency, therefore, may be tuned electrically with a gate. So-called ``gatemons'' benefit from the fact that the transmon shunt capacitor and circuit nonlinearity can be combined into a single planar element without the use of a potentially lossy oxide layer. Indeed, voltage-gated quantum circuits have been frequency-tuned by GHz in graphene junctions \cite{oliver_2019, renard1_2022, mandar_2022} and semiconductor weak links. \cite{marcus1_2016, petersson_2018, huo_2023, petersson_2022, lu_2023, higginbotham_2023}

In this paper, we show an alternative \textit{in situ} method to tune $I_c$ across an SNS junction. By driving supercurrent through a pair of superconducting leads coupled to conventional normal metal, we modify the SNS JJ critical current in a four-terminal Josephson junction (4TJJ) and in a two-junction SQUID loop. We find that $I_c$ is suppressed monotonically with supercurrent in both geometries, and we measure non-sinusoidal periodic oscillations of the SQUID critical current as a function external flux. Our method shows that, rather than inducing magnetic flux \textit{via} current drive, a current bias can directly modulate $E_J$ in a single SNS junction. This approach may offer more precise control over the frequency response of a single junction with reduced sensitivity to noise. 

Current-biased multiterminal JJs is a well-studied field. Specifically, multiterminal JJs have garnered attention for their correlated Josephson effects analyzed from the geometry of their critical current contours (CCCs). \cite{manucharyan_2020, finkelstein_2019, finkelstein_2021, pribiag_2020} This technique typically involves mapping the resistance across a particular junction as a function of bias across two or more junctions coupled to the same weak link. Correlated transport phenomena between sets of superconducting leads are then described by the overall shape of the CCC. Specifically, contour asymmetries have been attributed to the Josephson diode effect \cite{finkelstein_2023, pribiag_2023} and supercurrent-enhanced centrifugal features have been associated with Josephson phase synchronization between junctions \cite{finkelstein_2022} and correlated four-fermion transport across more than one pair of leads. \cite{kim_2022} Furthermore, in the clean limit, the CCC provides information about the multiterminal Andreev spectrum suggested to host non-local Cooper quartet supercurrents. \cite{melin_2023, pribiag_2024, kayyahla_2023} While a variety of rich physics have been predicted to exist in multiterminal JJs, little attention has been paid to how one may utilize correlated Josephson effects in superconducting quantum circuit architectures. 

Previous studies on multiterminal JJs typically use a ballistic normal metal like graphene as a weak link. We opt instead to use a diffusive normal metal which allows for more design flexibility. The devices here were fabricated by evaporating Au and Al in successive lithography steps on a Si substrate. In the multiterminal device geometry, we refer to the junction studied for its change in critical current as the ``sample'' junction, while the junction used to apply the bias is referred to as the ``control'' junction (borrowing language from V. Chandrasekhar \cite{chandra_2025}).  The sample junction leads lie vertically and the control junction leads attach to the sample junction at 45 degree angles. Current bias across the sample (control) junction we abbreviate $I_{smpl}$ ($I_{ctrl}$). See Figure 1a. 

To achieve critical currents on the order of $10 \; \mu$A at the temperatures $T = 300$ mK used in this experiment, steps were taken during the fabrication process to maximize the superconducting coupling energy across the junctions. See the Methods section for more details on the fabrication process. $I_c$ of a single diffusive SNS JJ with normal metal length $L$ decreases exponentially with the ratio $L/\xi_N$ where $\xi_{N} = \sqrt{\hbar D/k_B T}$ is the superconducting coherence length. \cite{belzig_1999} Here, $D = (1/3)v_f\ell$ is the electronic diffusion coefficient, $v_f$ is the Fermi velocity and $\ell$ is the elastic scattering mean free path of the electrons in the normal metal. We design $L$ long enough to attach the additional Au wires needed for the measurement while ensuring $L < L_\phi$, the normal electron phase coherence length (discussed further in the Supporting Information). The addition of normal metal arms reduces the base critical current of a SNS JJ \cite{noh_2013} but does not pose a significant issue for the purposes of this experiment. For the devices studied here, we estimate $\xi_N \sim  0.54 \; \mu$m$/\sqrt{T}$ based on sheet resistance measurements on devices from an identical fabrication process. \cite{wisne_2024} \\

\noindent\textbf{Results and Discussion}\\

Taking into account all design considerations, the maximum $I_c$ at $T=300\;$mK for the sample and control junctions were measured to be $\sim 10\;\mu$A and $\sim7.5 \;\mu$A, which were extracted from the transition to finite voltages in Figures 1b and 1c, respectively. Here, the differential resistance of each JJ was measured as a function of dc current while the other pair of leads were floated.  

\begin{figure}
\includegraphics[width = 8 cm]{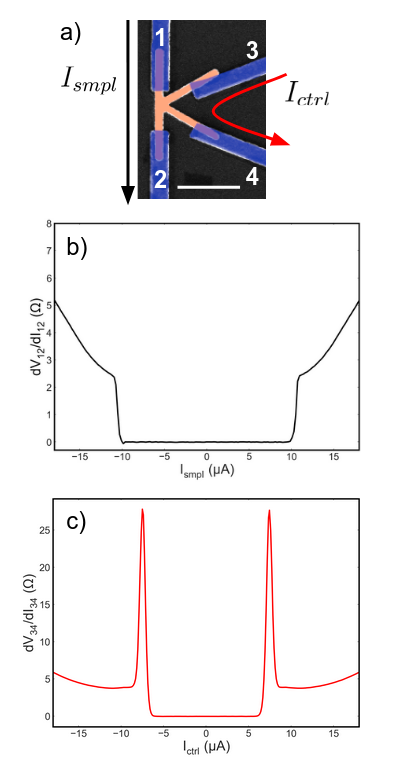}
\caption{\label{fig:svc} (a) False-colored SEM image of the 4TJJ showing both ``sample'' junction (leads 1 and 2) and ``control'' junction (leads 3 and 4). Blue is aluminum, orange is gold. Arrow indicates direction of dc bias current. Scale bar is $1 \; \mu$m. (b) Differential resistance $dV_{12}/dI_{12}$ measured across sample junction as a function of current bias $I_{smpl}$. (c) Differential resistance $dV_{34}/dI_{34}$ measured across control junction as a function of current bias $I_{ctrl}$. Smaller $I_c$ and large resistive peaks at the transition on control junction are due to the longer $L$ between superconducting electrodes.}
\end{figure}

The main goal of this experiment is to understand how the critical current of the sample junction changes with supercurrent across the control junction. To achieve this, the control junction was biased using a floating current source to limit the device to a single grounding point. Figure 2a shows a schematic of the measurement circuit along with the differential resistance $dV_{12}/dI_{12}$ of the sample junction at increments of the control junction bias $I_{ctrl}$. A voltage summer combined a $101$ Hz ac excitation with a dc voltage bias and the signal was converted into current by a home-made current source before being sent to ground through the sample junction. A relatively large $1 \;\mu$A ac excitation was used to broaden the resistive transition into the superconducting state in order to assist the locking circuit discussed later in the text.

\begin{figure}
\includegraphics[width = 8.5 cm]{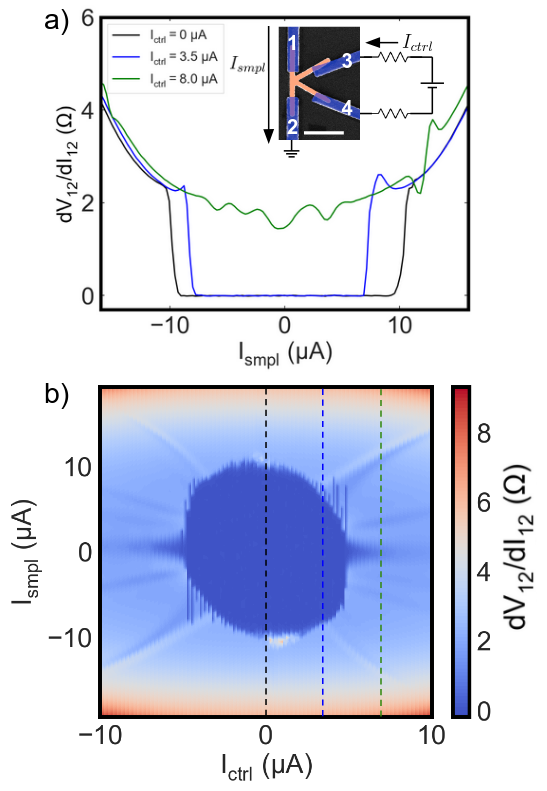}
\caption{\label{fig:4tjj} (a) Differential resistance as a function of current across the sample junction at three different values of control current. Inset: Circuit schematic showing the floating current source used for the control junction. Resistors are $100$ k$\Omega$. (b) Differential resistance heat map of sample junction as a function of sample and control bias currents. Perimeter of the zero-resistance area denotes the CCC as described in the introduction. Dotted vertical lines indicate traces in (a).}
\end{figure}

We discuss several important features from the resistive heat map in Figure 2b. First, and most relevant, applying supercurrent through the control junction substantially reduces the critical current of the sample junction. For example, driving $4\;\mu$A of supercurrent reduces the sample $I_c$ by half. Further increasing the bias of the control junction destroys the Josephson coupling across the sample junction. Beyond $I_c$ of the control junction, the $dV/dI$ heat map of the sample junction transitions into a mostly resistive state yet retains centrifugal streaking features that persist to regions of higher bias. The green trace in Figure 2a highlights a cross section of these features where transport across the control and sample junction is dominated by quasiparticle current. Importantly, these characteristics are absent from Figure 1 where only one junction is measured at a time. This suggests that the high bias features originate from cross-coupling between superconducting leads at different voltages. Similar streaking features observed in graphene-based multiterminal JJs have been attributed to non-local Andreev processes between more than two superconducting leads. \cite{kim_2022}  Given the diffusive nature of the normal metal in our devices, these streaks are unlikely to arise from bound state resonances. Instead, they likely indicate enhanced supercurrent flow when the voltage at one superconducting control terminal is commensurate with the finite potential across the sample junction as noted by Nowak \textit{et al}. \cite{akhmerov_2019}  While interesting, features that extend beyond the zero-resistance region of Figure 2 are not the focus of this paper. The main point is the dependence of the zero-resistance area on the control junction bias current which shows that $I_c$ of the sample junction depends on the amount of supercurrent passing through a shared weak link. 

Another way to observe $I_c$ dependence on supercurrent is in a SQUID configuration. We measure a device where two SNS JJs are connected in parallel to form a loop through which magnetic flux may be threaded.  Figure 3a details such a device. To map $I_c$ as a function of both flux and supercurrent, we establish a current bias feedback circuit locked to $I_c$ of the SQUID. A home-made PID controller is locked to the resistive peak corresponding to the transition out of the superconducting state as shown in Figure 3b. The output of the PID is combined with the ac excitation amplitude $I_{ac}$ and a dc offset current $I_{dc}$ used to initialize the SQUID at the critical current $I_c = I_{dc} + I_{PID}$. When properly locked, the PID output $I_{PID}$ adjusts to show the variation in critical current caused by flux and supercurrent bias. 

Let us first discuss how the total SQUID $I_c$ is affected by supercurrent bias $I_{ctrl}$. Due to the additional leads on only one side of the SQUID, the critical current $I_{c2}$ of the right JJ is expected to be smaller than the critical current $I_{c1}$ of the left JJ. The total field-dependent critical current of the SQUID, then, will oscillate between $I_{c1} + I_{c2}$ and $I_{c1} - I_{c2}$ as seen in Figure 3c. Since the SQUID JJs act in parallel, $I_{ctrl}$ influences both $I_{c1}$ and $I_{c2}$. We therefore expect the average value of the total SQUID $I_c$ and amplitude of oscillations to decrease with $I_{ctrl}$. Indeed, this behavior can be readily observed in the purple and blue traces of Figure 3c.

\begin{figure*}
\includegraphics[width = 18 cm]{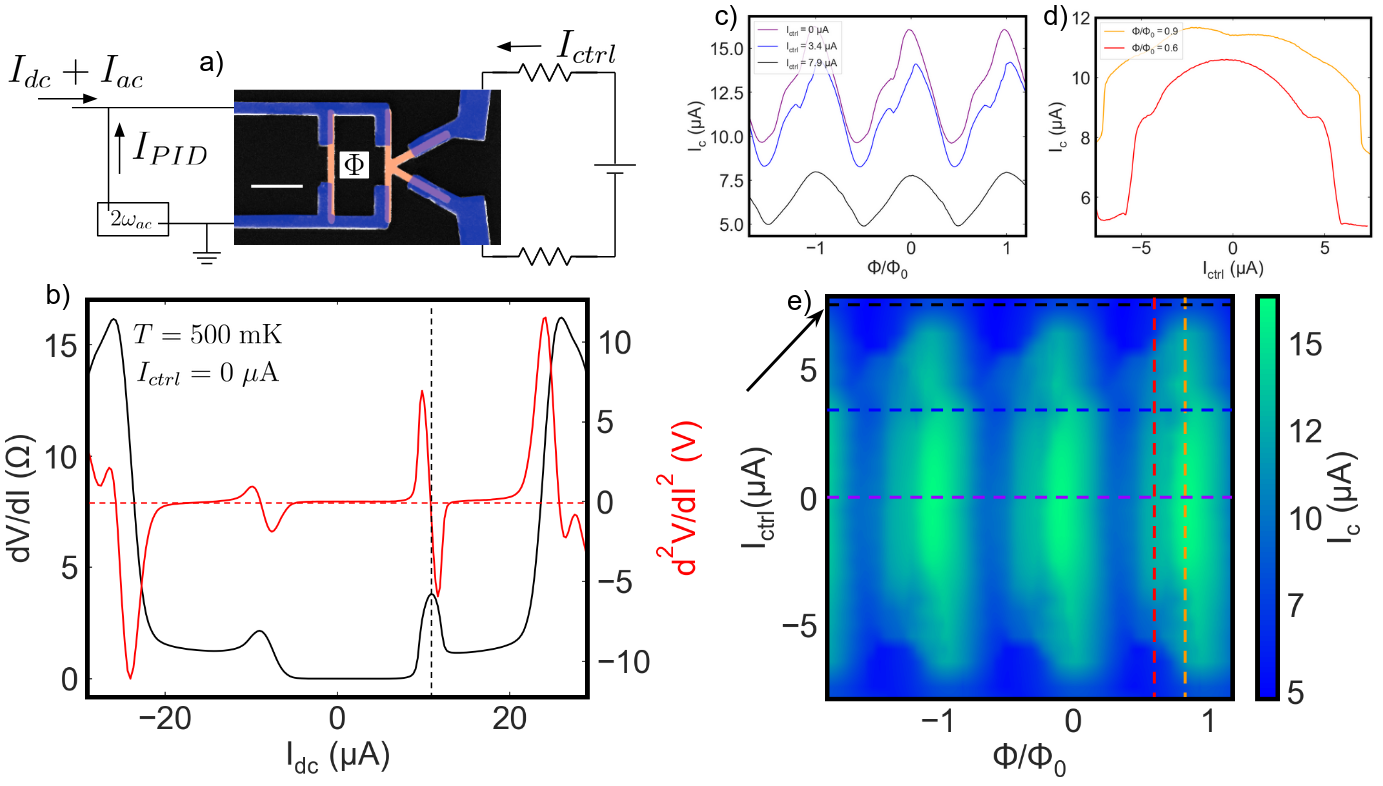}
\caption{\label{fig:ld} (a) Schematic of the measurement on the SQUID loop device. Scale bar represents $1\;\mu$m. The area of the loop is $\sim0.55\;\mu\text{m}^2$. The feedback circuit locks to the $2\omega_{ac}$ component of the second derivative of the ac voltage signal shown in (b). (b) The first (black) and second (red) derivative of $V(I)$ across the SQUID as a function of dc current captured at $500$ mK and zero control junction bias. PID setpoint indicated by red dotted line. (c) (d) Example traces of measurement locked to $I_c$ as a function of flux (supercurrent bias). (e) Full critical current heat map captured by sweeping external field and incrementally stepping $I_{ctrl}$. Dotted lines indicate traces in (c) and (d). Black arrow leads the eye to the trace in black.}
\end{figure*}

The complete SQUID $I_c$ dependence on $I_{ctrl}$ can be seen from the Figure 3e. We note several important features from this heat map. First, $I_c$ clearly oscillates with a fundamental period corresponding to the superconducting flux quantum $\Phi_0 =h/2e$ across all $I_{ctrl}$ as one might expect from the supercurrent transport across the device. Second, the current-phase relation of the multiterminal JJ is observed to be non-sinusoidal. At $I_{ctrl} = 0 \; \mu$A, $I_c$ of the SQUID is skewed with respect to flux. There appears to be an indication of a higher harmonic that develops as a function of control bias (see the Supporting Information). As $I_{ctrl}$ is increased, a phase difference $\phi_{ctrl}$ develops across the control junction. Recent theory predicts a phase difference across a SNS junction coupled to a separately-biased junction can modify the current-phase relation of the second junction. \cite{chandra_2025} Indeed, secondary features within the sinusoidal SQUID critical current oscillations appear to become more prominent at $I_{ctrl} \sim 3 -4\;\mu$A until completely vanishing when $\phi_{ctrl}$ appears to saturate. At this point, further increasing $I_{ctrl}$ has little effect on the current-phase relation of the SQUID.

The implementation of a flux-biased SQUID effectively enables configuration of the SNS JJ at various Josephson energies. Ramping the flux coupled to the loop past $\Phi = 0.5\Phi_0$, for example, causes the circulating current in the SQUID to switch directions, modifying the maximum critical current and Josephson energy of the device. Figure 3d shows two traces of $I_c$ as a function of $I_{ctrl}$ at various values of persistent flux $\Phi$. Both curves remain singularly monotonic, similar to the 4TJJ device. 

Also interesting in Figure 3d is the sharp transition to smaller $I_c$ in both the red and yellow traces. This is due to the majority of the supercurrent routing through the left JJ at particular values of $I_{ctrl}$ (see the Supporting Information for more detail). As evidenced by the black curve in Figure 3c, however, the SQUID retains coherent $h/2e$ oscillations above these transitions. We surmise that supercurrent preferably travels through the left JJ due to the modified quasiparticle density of states at the junction center resulting from a phase across the control junction.\cite{chandra_2025} Regardless, for values of $I_{ctrl}$ before the transition, the total SQUID $I_c$ is modulated by roughly 20\%, demonstrating that the Josephson energy of the two-junction loop is sensitive to supercurrent flowing through either branch, regardless of the effective initial Josephson energy of the device.

In conclusion, we study the critical current dependence on supercurrent in a 4-terminal SNS JJ in a standalone and SQUID loop geometry. We find a monotonic reduction in $I_c$ with incrementing supercurrent bias. In a SQUID configuration, we measure a non-sinusoidal current-phase relation by locking a feedback circuit to $I_c$. We find that the total Josephson energy of the SQUID can be reduced by approximately $20\%$ \textit{in situ} using a non-dissipative supercurrent drive. Our measurement suggests a possibility to augment the frequency response of quantum circuits fabricated from single normal metal JJs without the need to implement a two-junction loop which exposes the circuit to flux noise. \\

\noindent\textbf{Methods}\\

Devices included in this study were fabricated using e-beam lithography on Si substrates with a 1 $\mu$m SiO$_2$ insulating layer. All features were patterned in MMA/PMMA bilayers using a Tescan MIRA 4 electron microscope onto which ultra pure Au and Al films were deposited in an Edwards thermal evaporator used exclusively for 99.999\% (5N) Au and Al source material. Separate lithographic steps were used for the Au and Al films. Prior to the Au deposition, an \textit{in situ} O$_2^+$ plasma etch was used to clean the substrate.

The quality of the interfaces between the Au and Al wires also heavily influences the strength of the superconducting proximity effect and thus $I_c$. The proximity effect is improved by designing significant overlap between the Au and Al layers (see Figure 1a). Ar$^+$ plasma etching was used to clean the surface of the Au wires immediately prior to Al deposition. Additionally, the devices were cooled to cryogenic temperature within three hours of depositing Al to mitigate the formation of an intermetallic compound known to decrease Au/Al interface quality. \cite{chien_thesis} \\

\noindent\textbf{Acknowledgements}\\

This research was conducted with support from the National Science Foundation under grant No. DMR-2303536. \\

\noindent\textit{Supporting Information}: Discussion of relevant SNS junction length scales, and additional analysis of SQUID data including critical current dependence and non-sinusoidal current phase relation.

\bibliography{cc}

\end{document}